\documentclass[journal]{IEEEtran}
\usepackage{units}
\usepackage{ifpdf}
\usepackage{cite}
\ifCLASSINFOpdf
 \usepackage[pdftex]{graphicx}
\else
 \usepackage[dvips]{graphicx}
\fi
\usepackage[cmex10]{amsmath}
\usepackage{array}
\usepackage{mdwtab}
\usepackage[font=footnotesize]{subfig}
\usepackage{url}

\providecommand{\sub}[1]{_{\rm #1}}
\providecommand{\abl}[2]{\frac{{\rm d} #1}{{\rm d} #2}}  
\providecommand{\ablpart}[2]{\frac{\partial #1}{\partial #2}}  
\providecommand{\twoCases}[4]{
  \left\{ 
    \begin{array}{ll} 
      #1 & #2 \\
      #3 & #4 
    \end{array} 
  \right.
}

\begin{document}

\title{Automatic and efficient driving strategies while approaching a traffic light}

\author{Martin~Treiber\thanks{M.~Treiber is with the Department of Transport and Traffic Sciences, 
Technische Universit\"at Dresden, W\"urzburger Str. 35, 01187 Dresden, Germany. 
e-mail: treiber@vwi.tu-dresden.de}, 
and Arne~Kesting\thanks{A.~Kesting is with TomTom Development Germany, An den Treptowers~1, 
12435 Berlin, Germany. e-mail: mail@akesting.de}
}%


\maketitle

\begin{abstract}
Vehicle-infrastructure communication opens up new ways to improve
traffic flow efficiency at signalized 
intersections. In this study, we assume that equipped vehicles can
obtain information about switching times of relevant traffic lights in
advance. This information is used to improve traffic flow by the
strategies ``early braking'', ``anticipative start'', and
``flying start''. The strategies can be implemented in driver-information mode, or in automatic mode by an Adaptive
Cruise Controller~(ACC). Quality criteria include cycle-averaged
capacity, driving comfort, fuel 
consumption, travel time, and the number of stops. 
By means of simulation, we
investigate the isolated strategies and the complex interactions
between the
strategies and between equipped and non-equipped vehicles. As
universal approach to assess equipment level effects we propose
relative performance indexes and found, at a maximum speed of~\unit[50]{km/h}, improvements of about 15\% for
the number of stops and about~\unit[4]{\%} for the other
criteria. All figures double when increasing the maximum speed to~\unit[70]{km/h}. 
\end{abstract}


\IEEEpeerreviewmaketitle

\section{\label{sec:intro}Introduction}
%
\IEEEPARstart{I}{ndividual} vehicle-to-vehicle and vehicle-to-infrastructure
communication, commonly refer\-red to as V2X, are
novel components of intelligent traffic systems
(ITS) \cite{papadimitratos2009vehicular,hartenstein2010vanet}. Besides more traditional ITS applications
such as variable speed limits on freeways~\cite{papageorgiou2008effects} or
traffic-dependent signalization~\cite{SCOOT,SCATS}, V2X promises new
applications to make traffic flow more efficient or driving more
comfortable and economic~\cite{bishop2005intelligent}.
While there are many investigations
focussing on technical issues such as connectivity given a certain hop
strategy, communication
range, and percentage of equipped
vehicles (penetration rate), e.g., \cite{kestingIvc2010,jin2010analytical,thiemann-IVC-PRE08}, few papers
have investigated actual strategies to improve traffic flow
characteristics~\cite{van2006impact,Arne-ACC-TRC}. On 
freeways, a jam-warning system based on communications to and from
road-side units (RSUs) has been proposed~\cite{Kranke-Fisita2008}. Furthermore, a
traffic-efficient adaptive-cruise control (ACC) has been proposed
which relies on V2X communication to determine the local traffic
situation influencing, in turn, the ACC
parameterization~\cite{kesting-acc-roysoc}. Regarding city traffic,
V2X have been investigated as early as
1991~\cite{PrometheusDrive1991}. Nowadays, V2X is an active research
field with
several research initiatives such as 
Compass~\cite{Compass}. However, most of
these initiatives 
focussed on safety and routing information without explicitly
treating any interactions with traffic lights. The investigations which are
arguably most related to our work are that in the research projects
Travolution~\cite{braun2008travolution} and the sub-project ``smart intersection''
of the multi-institutional research initiative ``UR:BAN''~\cite{URBAN} combining 
traffic-adaptive signalization (V2X) with driver information on the future state
of the signals (X2V). Adapting the driving behavior based on this
information has the potential to
reduce fuel consumption~\cite{tielert2010impact}. However, in the 
thesis~\cite{Otto2011kooperative}, a possible destructive
interplay between V2X and X2V has been found.

In this contribution, we focus on city traffic at signalized
intersections and investigate a set of strategies that is complementary to
the self-controlled signal control strategy of L\"ammer and Helbing~\cite{lammer2008SelfControl}: While, in the
latter, the traffic lights ``know'' the future traffic, we assume that
equipped vehicles know the future states of the next traffic
light by X2V communication. In principle, the proposed \textit{traffic-light assistant} (TLA) can
operate in the information-based manual mode as in
~\cite{Otto2011kooperative,iglesias2008i2v}, or in the ACC-based 
automatic mode on which we will focus in this work.

In the following Section~\ref{sec:methodology}, we lay out the methodology of this
simulation-based study and define the objectives. Section~\ref{sec:strategies}
presents and analyzes the actual strategies ``economic approach'',
``anticipative start'', ``flying start'', and their interplay.
In the concluding Section~\ref{sec:diss}, we discuss the results and point at
conditions for implementing the strategies in an actual TLA.

\section{\label{sec:methodology}Methodology}

\subsection{\label{sec:idm}Car-Following Model}
%
In order to get valid results, the underlying car-following model must
be (i)~sufficiently realistic to represent ACC driving in the
automatic mode of the TLA, (ii)~simple enough for
calibration, and (iii)~intuitive enough to readily implement the new
strategies by re-parameterizing or augmenting the model. We apply the
``Improved Intelligent-Driver Model''
(IIDM) as described in Chapter~11 of the
book~\cite{TreiberKesting-Book}.\footnote{This chapter is available
 for free at\\ \texttt{www.traffic-flow-dynamics.org}} As the original Intelligent-Driver
Model (IDM)~\cite{Opus}, it is a
time-continuous car-following model with a smooth acceleration
characteristics. Assuming speeds $v$ not exceeding the desired speed
$v_0$, its acceleration equation as a
function of the (bumper-to-bumper) gap $s$, the own speed $v$ and the speed $v_l$ of the
leader reads
\begin{equation}
\label{IIDM}
\abl{v}{t}
=\twoCases{(1-z^2)\, a}
{\quad z=\frac{s^*(v,v_l)}{s}\ge 1,}
{ \left(1-z^{\frac{2a}{a\sub{free}}}\right)\, a\sub{free}}
{\quad \text{otherwise,}}
\end{equation}
where the expressions for the desired dynamic gap $s^*(v,v_l)$ and the
free-flow acceleration $a\sub{free}(v)$ are the same as that of the
IDM,
\begin{eqnarray}
\label{sstar}
s^*(v,v_l) &=& s_0
+\max\left[vT+\frac{v(v-v_l)}{2\sqrt{ab}}, \, 0\right],\\
a\sub{free}(v) &=& a
\left[1-\left(\frac{v}{v_0}\right)^{\delta}\right].
\end{eqnarray}
The IIDM has the same parameter set as the IDM:
desired speed $v_0$, 
desired time gap $T$, minimum space gap $s_0$, desired
acceleration $a$, and desired deceleration $b$. However, it resolves
two issues of the basic IDM when using it as an ACC acceleration
controller: (i)~the IIDM time-gap parameter $T$ describes exactly the time
gap in steady-state car-following situations while the actual IDM
steady-state time gaps are somewhat larger~\cite{Opus}, (ii)~a platoon of
vehicle-drivers with same desired speed $v_0$ will not disperse over time
as would be the case for the IDM. 
 
By describing the vehicle motion with a time-continuous car-following
model, we have neglected the in-vehicle
control path since such models implicitly
reflect an acceleration response time of zero. 
It might be necessary to explicitly model vehicle
responses by explicit delay and PI elements when actually
deploying such a system.

\begin{figure}[!ht]
 \begin{center}
  \includegraphics[width=0.95\linewidth]{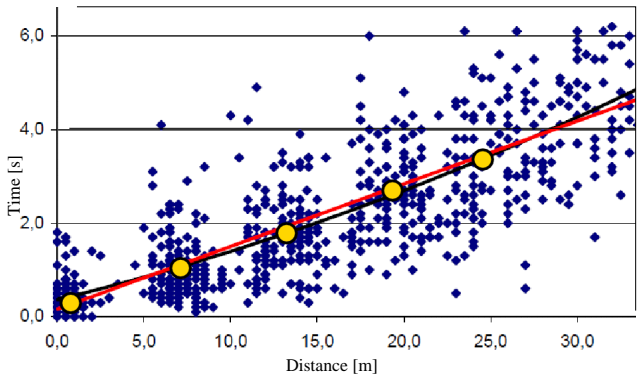}
 \end{center}
 \caption{\label{fig:calibr}Calibration of the microscopic model with respect to the
  starting times and positions of a queue of waiting vehicles
  relative to the begin of the green phase (solid circles). Data are of the
  measurements in Ref.~\cite{Kuecking}.}
\end{figure}

\subsection{\label{sec:calibr}Calibration}

Since we will
investigate platoons travelling from traffic light to traffic light,
the acceleration model parameters $v_0$, $T$, $s_0$, $a$, and $b$ and
the vehicle length $l\sub{veh}$ (including their variances) should be
calibrated to data of starting and stopping situations.

\begin{figure}[!ht]
 \begin{center}
  \includegraphics[width=\linewidth]{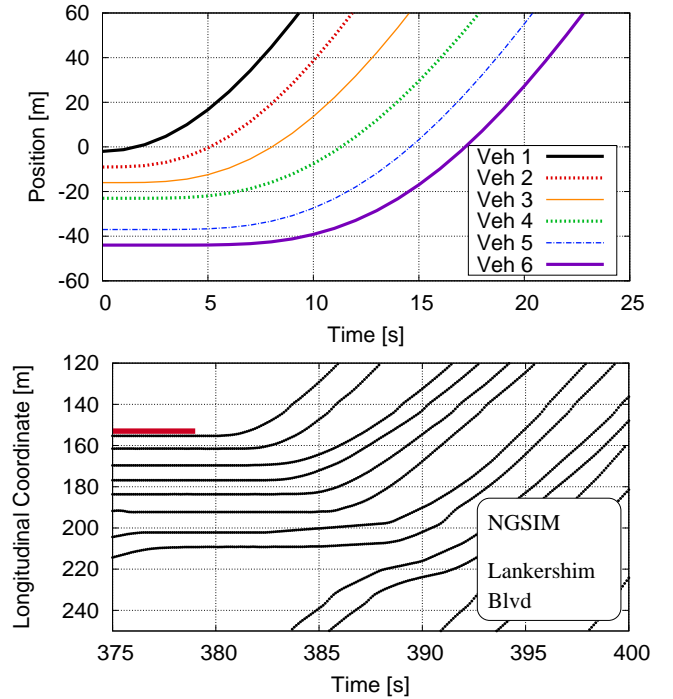}
 \end{center}
 \caption{\label{fig:traj}Trajectories of the start of the simulated platoon in
  comparison with trajectories from the 
  NGSIM initiative~\cite{NGSIM}.  }
\end{figure}

For calibrating $a$, $T$, and
the combination $l\sub{eff}=l\sub{veh}+s_0$ (effective vehicle
length), we use the empirical
results of K\"ucking~\cite{Kuecking} taken at three intersections in
the city of Hannover, Germany. There,
the ``blocking time'' of the $n^\text{th}$ vehicle of a waiting
queue (the time interval this vehicle remains stopped after the light
has turned green) has been measured vs. the distance of this vehicle to the stopping
line of the traffic light. Figure~\ref{fig:calibr} reproduces these
data together with the simulation results (orange bullets) for the
calibrated parameters $l\sub{eff}=\unit[6.5]{m}$,
$a=\unit[1.5]{m/s^2}$, and $T=\unit[1.2]{s}$ assuming 
identical vehicle-driver units. As shown in Fig.~\ref{fig:traj} for the start, the resulting
trajectories are comparable to observed trajectories on the Lankershim
Blvd as obtained from the NGSIM
initiative~\cite{NGSIM}. Further simulations with heterogeneous
drivers and vehicles reveal that independently and uniformly distributed values for
$l\sub{eff}$, $T$ and $a$ with standard deviations of the order of
\unit[30]{\%} of the respective expectation value can reproduce the
observed data scatter and its increase with the vehicle position (for
positions $n=5$ and higher, the scattering does no longer allow to
identify $n$). Moreover, since trucks are excluded from the
measurements, it is reasonable to assume that the observed cars have
an average length of~\unit[4.5]{m} resulting in an expectation value
$s_0=\unit[2]{m}$ for $s_0$.

For estimating the comfortable deceleration, the approach to a red
traffic light is relevant. Trajectories of the Lankershim data set
of the NGSIM data~\cite{NGSIM} including such situations
(Fig.~\ref{fig:traj}) indicate that
typical decelerations are comparable to typical accelerations. We assume
$b=\unit[2]{m/s^2}$~\cite{viti2010microscopic}.
Finally, for the desired speed, we assumed a fixed value of
$v_0=\unit[50]{km/h}$ representing the usual inner-city speed limit in Germany.

\begin{figure}[!ht]
 \begin{center}
  \includegraphics[width=\linewidth]{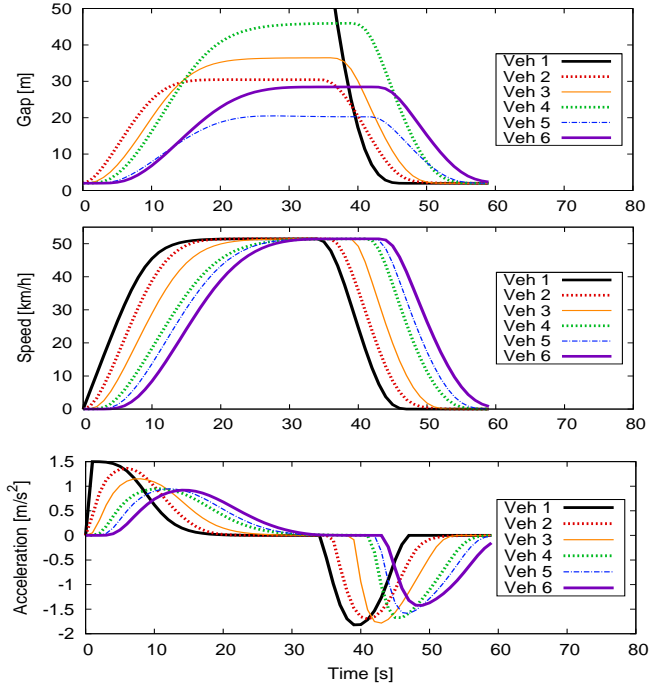}
 \end{center}
 \caption{\label{fig:reference}Start and stop of the simulated platoon of heterogeneous
  vehicle-drivers in the reference case. }
\end{figure}

\subsection{\label{sec:sim}Simulation}

While the parameters clearly are distributed due to inter-vehicle and inter-driver
variations, it is nevertheless necessary to use the same vehicle
population for all the following simulation experiments. Specifically,
we use following sequence of four vehicle-driver combinations: 1. average driver
(expectation values for the parameters),
2. agile driver ($a$ increased to $\unit[2]{m/s^2}$, $T=\unit[1.8]{s}$),
3. less agile but anticipative driver ($a$ and $b$ decreased to
$\unit[1.2]{m/s^2}$ and $\unit[1]{m/s^2}$, respectively), 
and 4. a truck
($l\sub{veh}=\unit[12]{m}$, $T=\unit[1.7]{s}$, and
$a=b=\unit[1]{m/s^2}$).
 If necessary, this sequence is
 repeated. The Figures~\ref{fig:traj} and~\ref{fig:reference} show the simulation result for
 the start-and-stop reference scenario against which the strategies of
 the traffic light assistant will be tested in the next section.

\subsection{Traffic Flow Metrics}

In the ideal case, the TLA reduces the travel time of the equipped and
the other vehicles, increases driving comfort and traffic flow
efficiency, and reduces fuel
consumption~\cite{tielert2010impact}. To assess travel
time, we use the average speed of a vehicle, or average over all vehicles during
the complete simulation run. As proxy for the driving comfort, we take
the number
of stops during one simulation, or, equivalently, the fraction of
stopped vehicles. Traffic flow efficiency is equivalent to the
cycle-averaged dynamic capacity, i.e., the average number of vehicles
passing a traffic light per cycle in congested conditions in the
absence of gridlocks. Finally, we determine the fuel consumption
 by a physics-based modal
consumption model as described in Chapter~20.4 of Ref.~\cite{TreiberKesting-Book}. 
Such models take the simulated
trajectories and some vehicle attributes 
as input and return the instantaneous consumption rate
and the total consumption of a given vehicle. To be specific, we
assume a mid-size car with following attributes: 
Characteristic map of a \unit[118]{kW} gasoline engine as in Fig.~20.4
of~\cite{TreiberKesting-Book}, idling
power $P_0=\unit[3]{kW}$, total mass $m=\unit[1500]{kg}$,
friction coefficient $\mu=0.015$, air-drag coefficient $c_d=0.32$,
frontal cross-section $A=\unit[2]{m^2}$, a dynamic tire radius
$r\sub{dyn}=\unit[0.286]{m}$. Furthermore, we assume a five-gear transmission with
transmission ratios of 13.90, 7.80, 5.25, 3.79, and 3.09,
respectively, and choose
the most economic gear for a given driving mode characterized by $v$
and $\abl{v}{t}$. The engine power management includes overrun-fuel
cutoff, idling when the vehicle is stopped, and no energy recuperation
during braking. 

\section{\label{sec:strategies}Strategies of the Traffic Light Assistant and their Simulation}
\begin{figure}[!ht]
 \begin{center}
  \includegraphics[width=0.85\linewidth]{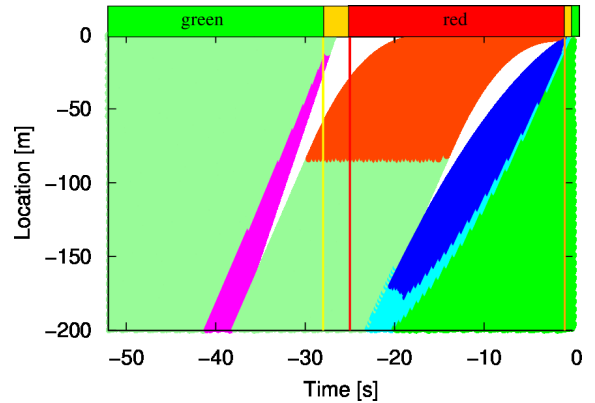}
 \end{center}
 \caption{\label{fig:KOLINE}Approach situations relative to the phases of
  the traffic light (stopping line at $x=0$). For the color coding, see
  the main text.}
\end{figure}

%
The appropriate TLA strategy depends essentially on the arrival time
at the next traffic light
relative to its phases. Depending on the spatiotemporal position, we
distinguish following approaching situations
(cf. Fig.~\ref{fig:KOLINE}):

\begin{itemize}
\item A stop is unavoidable and the vehicles are sufficiently near the
 signal to initiate braking (red spatiotemporal region
 of Fig.~\ref{fig:KOLINE}),
\item anticipative start compensating for the reaction time of the
 first vehicle (the last two seconds of the red area),
\item flying start realized by anticipative engine braking (blue) or proper
 braking (turquoise),
\item free passage or sufficiently away from the intersection, so no
 action is necessary (green), and 
\item temporary ``boost'' to catch the last part of the green phase
 (violet region). 
\end{itemize}

Nothing can be done in the situation of a free passage. Moreover, the
``boost'' strategy implies temporarily exceeding the speed
limit which we will not pursue in this contribution. In the following,
we 
will develop and simulate the remaining three strategies ``approach to
a stop'', ``anticipative start'', and ``flying start''. Generalizing the above sketch, we
will also investigate how other (equipped or
non-equipped) vehicles will affect the strategies. Furthermore, by a
complex simulation over several cycles, we investigate
any (positive or negative) interactions between the strategies and
between equipped and non-equipped vehicles.

\subsection{\label{sec:approachStop}Approach to a Stop}

In certain situations, a stop behind a red light or a waiting queue is
unavoidable. This situation is true if (i)~extrapolated
constant-speed arrival occurs during a red phase, and (ii)~the
``flying-start'' strategy of Sect.~\ref{sec:flyingStart} would produce
minimum speeds below a certain threshold which we assumed to be
$v^\text{flying}\sub{min}=\unit[10]{km/h}$.\footnote{No engine braking
 is feasible below this speed.} Notice that this scenario
may also apply for approaching green traffic lights if the car cannot make it
to the traffic light before switching time: In such a situation, drivers of non-equipped
cars would just go ahead braking later and
necessarily harder.
While this situation is not relevant for improving flow efficiency, it is
nevertheless possible to reduce fuel consumption by early use of the engine
brake, i.e., early activation of the overrun cut-off.

In the car-following model, we implement this strategy by 
reducing the comfortable deceleration from $b=\unit[2]{m/s^2}$ to
$\unit[1]{m/s^2}$ (homogeneous driver-vehicle population), or by
\unit[50]{\%} for each vehicle (heterogeneous population). Reducing
the desired deceleration means earlier braking, in line with this strategy.

\begin{figure*}[!ht]
 \begin{center}
  \includegraphics[width=0.8\linewidth]{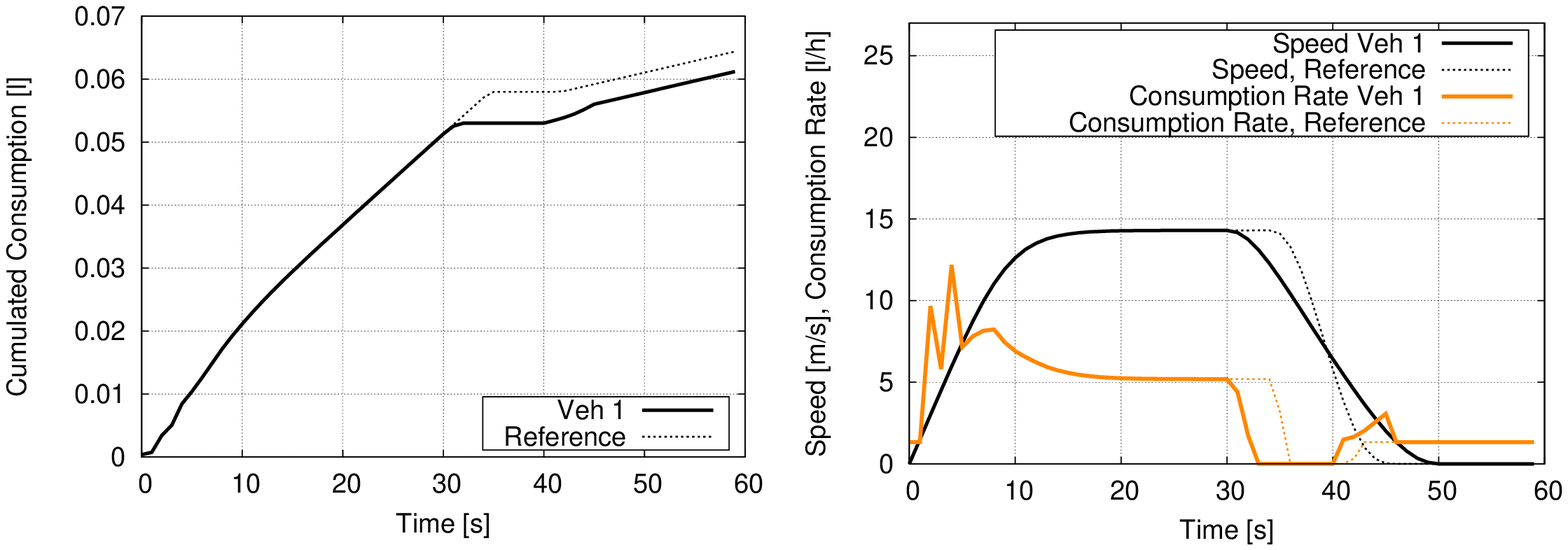}
  \includegraphics[width=0.8\linewidth]{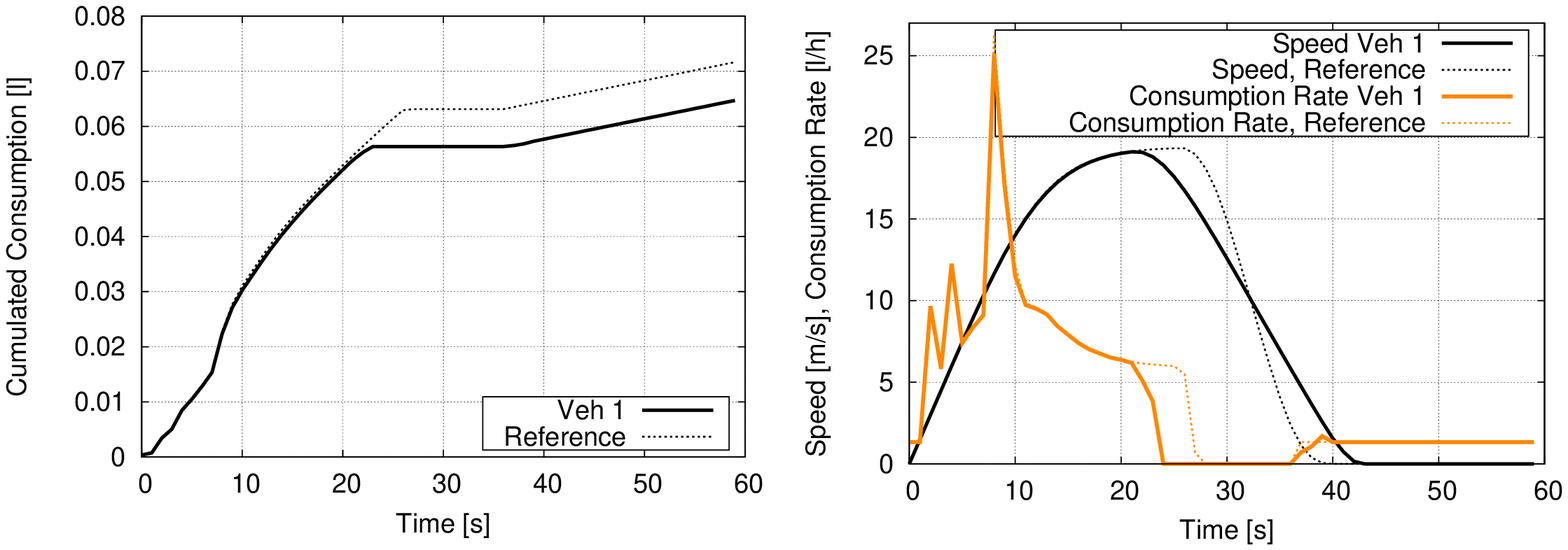}
 \end{center}
 \caption{\label{fig:approachStop}Fuel-saving approach to a waiting
  queue for speed limits of \unit[50]{km/h} (top) and \unit[70]{km/h} (bottom). 
  Left: cumulative consumption; right: speed profile and instantaneous consumption rate
  during the complete start-stop cycle.}
\end{figure*}

Figure~\ref{fig:approachStop} shows speed and consumption profiles for
an equipped vehicle (solid lines) vs. the reference (dotted). For a
speed limit of~\unit[50]{km/h}, the equipped vehicle
itself saves about \unit[3.5]{ml} of fuel (\unit[6]{\%} for the
complete start-stop cycle). The two next (non-equipped)
followers save about
\unit[3]{\%} and \unit[1]{\%}, respectively. For a limit of
\unit[70]{km/h}, the potential for saving is significantly higher.

\subsection{\label{sec:anticipativeStart}Anticipative Start}

\begin{figure*}[!ht]
 \begin{center}
  \includegraphics[width=0.8\linewidth]{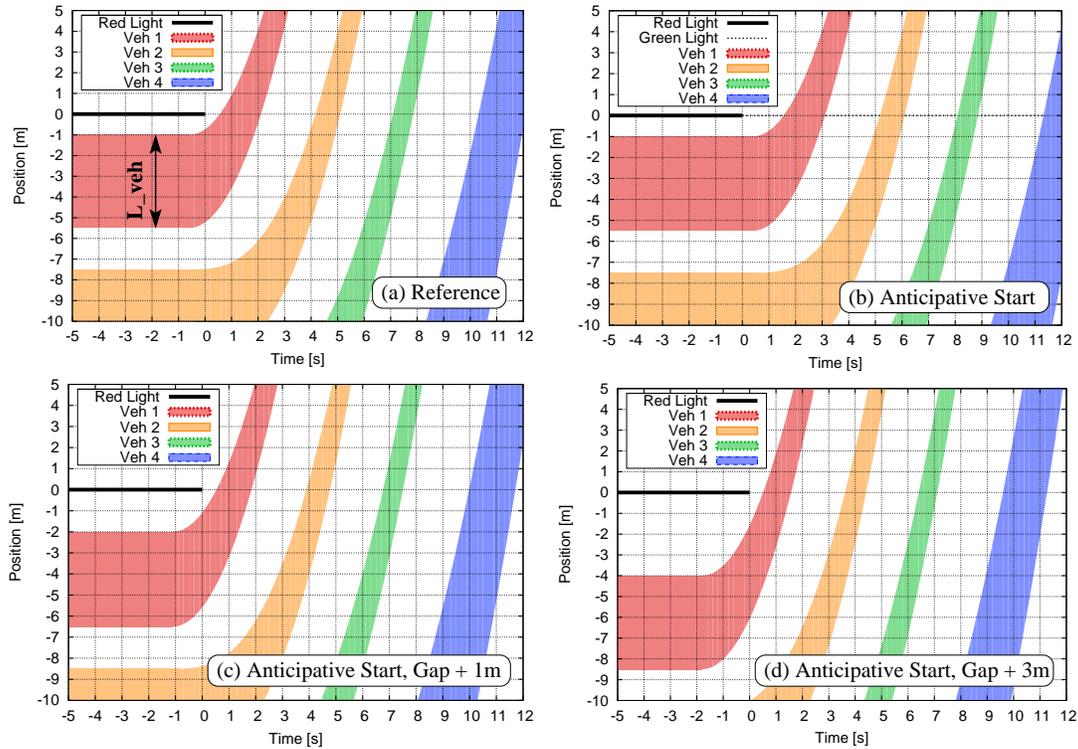}
 \end{center}
 \caption{\label{fig:stopStart}Start at green from the first position of a queue of
  waiting vehicles. (a)~reference; (b)~anticipative
  start; (c)~anticipative start plus \unit[1]{m} additional
  gap; (d)~anticipative start plus \unit[3]{m} additional
  gap.}
\end{figure*}

The rationale of the strategy of the anticipative start is to
compensate for the reaction time delay $\tau$.
Since the reaction time is only relevant
for the driver of the first vehicle in a queue, the anticipative-start
strategy is
restricted to this vehicle (see also the empirical
trajectory data, Fig.~\ref{fig:traj} bottom). In the reference case corresponding
to the calibrated parameters (Fig.~\ref{fig:stopStart}(a)), the front of
the first vehicle crosses the stopping line about \unit[1.5]{s} after
the change to green corresponding to $\tau\approx\unit[0.7]{s}$ (the
rest of the time is needed to move the first meter to the
stopping line). If this vehicle started one second earlier, i.e.,
before the switching to green
(Fig.~\ref{fig:stopStart}(b)), the situation is yet safe but an average
of $0.5$ additional vehicles can pass during one green phase assuming
an outflow of \unit[1800]{veh/h} after some vehicles. Considering
the $12$~vehicles that would pass in the reference scenario during the
\unit[30]{s} long green phase of the \unit[60]{s} cycle, this amounts,
on average, to an increase by \unit[4]{\%}.
An additional second can be
saved, allowing~$13$ instead of $12$~vehicles per green phase, if
the first vehicle stops \unit[4]{m} upstream of the stopping line
(instead of \unit[1]{m}) allowing an even earlier start without
compromising the safety
(Fig.~\ref{fig:stopStart}(d)). However, there are limits in terms of
acceptance and available space, so stopping \unit[2]{m} before the
stopping line (Fig.~\ref{fig:stopStart}(c)) is more realistic. In
effect, the latter strategy variants transform the anticipative start in a
``flying start'' which we will discuss now.

\subsection{\label{sec:flyingStart}Flying Start}

\begin{figure}[!ht]
 \begin{center}
  \includegraphics[width=\linewidth]{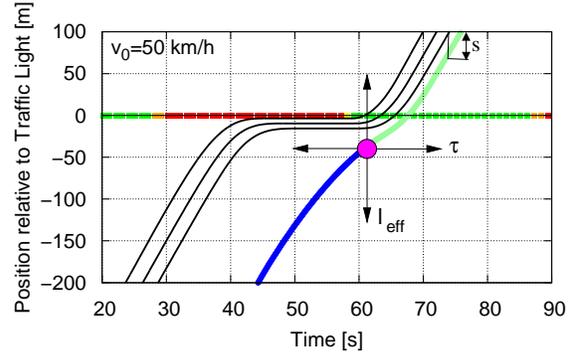}
 \end{center}
 \caption{\label{fig:flyingStart}Preemptive braking to avoid a stop: 
  Spatiotemporal target for the 4$^\text{th}$ vehicle (pink
  circle). The arrows indicate how the target changes when
  varying the reaction time $\tau$ or the
  effective length $l\sub{eff}$.}
 
\end{figure}

If, relative to the phases, a vehicle arrives later than in the
previous two situations but too early to have a free passage,
preemptive braking may avoid a stop or, at least,
increase the minimum speed during the approaching phase. As depicted
in Fig.~\ref{fig:flyingStart}, the strategy
consists in controlling the vehicle's ACC such that a certain
spatiotemporal \emph{target point} $(\Delta
 x,\Delta t$) relative to the stopping line and the switching time to
 green is reached. This point is determined such that a minimum of
 speed reduction is realized without impairing
 traffic efficiency by detaching this vehicle from the platoon of
 leaders. Generally, the braking is realized by
 engine braking modeled with a physics-based force model as
 described in Chapter~20.4 of Ref.~\cite{TreiberKesting-Book} and
 additional mild proper braking (deceleration 
 $\unit[1]{m/s^2}$) at the beginning of
 the deceleration whenever kinematically necessary.
 
From basic kinematic theory~\cite{Lighthill-W} and the properties of the IDM (Sect.~\ref{sec:idm}) it
follows that the propagation 
velocity $c$ of the \emph{positions} of the vehicles at the respective
starting times is
constant and given
by $c \approx
-l\sub{eff}/\tilde{T}$ where $\tilde{T}$ is of the order of the IDM
parameter $T$. Assuming a gap $s_0^*$ of the first waiting vehicle to
the stopping line and a reaction delay $\tau$ of its driver, the
estimated spatiotemporal \emph{starting point} of the 
$n^\text{th}$ vehicle reads
\begin{equation}
\label{targetPoint}
(\Delta x, \Delta t)=( s_0^*+[n-1]\, l\sub{eff}, \tau+[n-1] \, \tilde{T}).
\end{equation}
The points lie on a straight line which is consistent with observations (filled circles in
Fig.~\ref{fig:calibr}). While we assume that, by additional V2X communication from a
stationary detector to the vehicle, the equipped vehicle knows its order number $n$,
there are uncertainties in $\tau$, $l\sub{eff}$, and $T$ which depend on
unknown properties of the vehicles and drivers ahead. Furthermore, since the
strategy tries to avoid a stop, the 
\emph{target} point lies several meters upstream of and/or a few seconds
after the anticipated starting point. 

Is this strategy nevertheless robust? In order to assess this, we treat $\tau$ and
$l\sub{eff}$ (cf. Fig.~\ref{fig:flyingStart}) as free parameters of Eq.~\eqref{targetPoint} 
to be estimated and plot the performance metrics
spatial gap $s$ to the platoon (characterizing the dynamic capacity) and 
minimum speed $v\sub{min}$ (characterizing driving comfort) 
as a function of $\tau$ and $l\sub{eff}$. 

\begin{figure}[!ht]
 \begin{center}
  \includegraphics[width=0.8\linewidth]{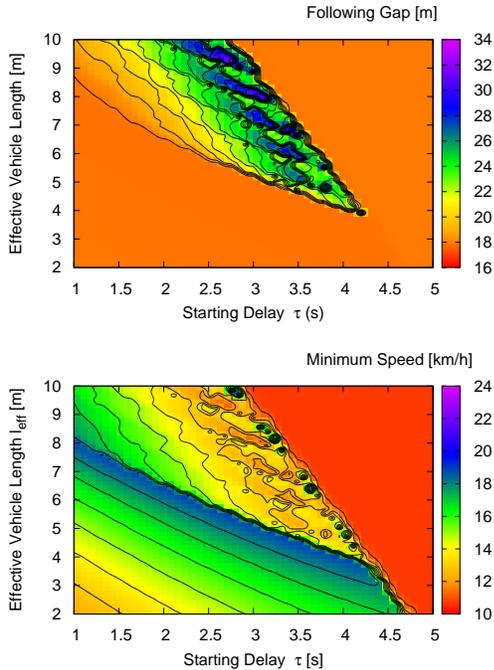}
 \end{center}
 \caption{\label{fig:flyingStartTauLeff}Robustness of the preemptive braking strategy for a maximum
  speed of \unit[50]{km/h}. Shown is its
  efficiency for the $3^\text{rd}$ vehicle in terms of
  the minimum speed during the approach (left) and the gap
  once this vehicle is \unit[50]{m} downstream of the traffic light
  (right).} 
 
\end{figure}

Figure~\ref{fig:flyingStartTauLeff} shows these metrics
for the $n=3^\text{rd}$ vehicle arriving at a timing such that
the minimum speed would be 
$v\sub{min}=\unit[10]{km/h}$ if this vehicle were not equipped. For the best
estimates (e.g., $l\sub{eff}=\unit[6.5]{m}$ and $\tau=\unit[2]{s}$), this minimum speed is nearly doubled without compromising
the capacity which would be indicated by an increased following
gap $s$. The simulations also show that estimation errors have one of three
consequences: (i)~if the queue length and
dissolution time are estimated too optimistically ($l\sub{eff}$ and
$\tau$ too small), there is still a
positive effect since the minimum speed is increased without
jeopardizing the efficiency; (ii)~if the queue is massively overestimated
($l\sub{eff}$ and $\tau$ significantly too large), the whole strategy is deemed unfeasible
and the approach reverts to that of non-equipped
vehicles; (iii)~if, however, the queue is only slightly overestimated,
the strategy kicks in ($v\sub{min}$ increases) but the capacity is
reduced since $s$ increases as well: the car does no longer catch the
platoon. A look at the parameter ranges (the plots range over factors of five in both
$\tau$ and $l\sub{eff}$) indicates that this strategy is
robust when erring on the optimistic side, if there is any doubt. 

Finally, we mention that counting errors (e.g. due to a vehicle changing
lanes when approaching a red traffic light meaning that this vehicle
has not passed the correct stationary detector) will lead to similar errors
for the estimated target point as above. Consequently, this strategy should be
robust with respect to counting errors as well.

\subsection{\label{sec:complex}Complex Simulation}

In the previous sections, we have investigated the different strategies
of the TLA in isolation. However, there are interactions. For example, the
optimal target point of the flying-start strategy is shifted backwards
in time when equipped leading vehicles apply the
anticipative-start strategy. Furthermore, the question remains if the
TLA remains effective if there is significant surrounding traffic (up
to the level of saturation) and whether the results are sensitive to the
order in which slow and fast, equipped and non-equipped vehicles
arrive.

We investigate this by complex simulations
of all strategies over several cycles where we vary, in each
simulation, the overall traffic 
demand (inflow) $Q\sub{in}$, and the penetration rate $p$ of equipped vehicles. Unlike
the simulations of single strategies, we allow for full stochasticity
in the vehicle composition. At inflow, we draw, for each new vehicle, the model parameters
from the independent uniform distributions specified in Sect.~\ref{sec:sim} and
assign, with a probability $p$, the property ``is equipped''.

\begin{figure}[!ht]
 \begin{center}
  \includegraphics[width=\linewidth]{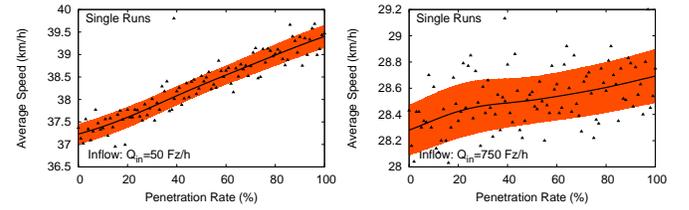}
 \end{center}
 \caption{\label{fig:complexRun50}Complex simulation of the overall effectiveness for all
  vehicles over several cycles in terms of the average speed for a maximum speed of
  \unit[50]{km/h}. See the main text for details.}
\end{figure}

\begin{figure}[!ht]
 \begin{center}
  \includegraphics[width=\linewidth]{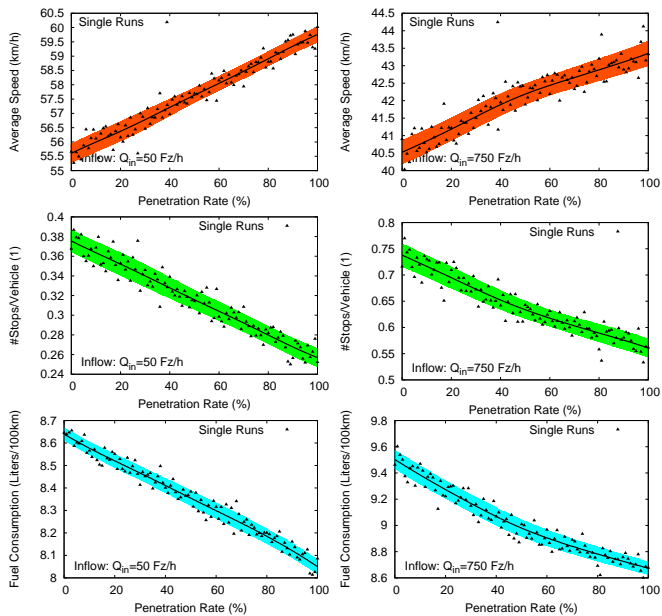}
 \end{center}
 \caption{\label{fig:complexRun70}Complex simulation of the overall effectiveness (as in
  Fig.~\ref{fig:complexRun50}) for the three criteria average
  speed, number of stops, and fuel consumption with assuming a maximum speed of
  \unit[70]{km/h}. Left column: low demand; right column: high
  demand near saturation.
}
\end{figure}

Figure~\ref{fig:complexRun50} shows the results for the performance
metrics ``average speed'' (which is related to the average travel
time) as a function of the penetration rate for
a small traffic demand (left) and near saturation
(right). Each symbol corresponds to a simulation for given values of
$Q\sub{in}$ and $p$. Due to the many stochastic factors and
interactions, we observe a wide scattering. Determining the local average
(solid lines) and $\pm 1 \sigma$
bands (colored areas) by kernel-based
linear regression (kernel width
\unit[15]{\%}), we nevertheless detect significant systematic
effects. For low traffic demand, we observe that travel times 
 are reduced by about \unit[4]{\%} when going from the
reference to $p=\unit[100]{\%}$
penetration. Similar figures apply to the fuel consumption. 
Furthermore, the effects essentially increase linearly with $p$, so the
\emph{relative performance indexes} $I\sub{T}$ and $I\sub{C}$ with respect to
travel time $T_t$ and fuel consumption $C$,
\begin{equation}
\label{perfIndex}
I\sub{T_t}=-\frac{1}{T_t}\ablpart{T_t}{p}, \quad
I\sub{C}=-\frac{1}{C}\ablpart{C}{p}
\end{equation}
are both constant and of the order of \unit[4]{\%}. The performance
index relative to the number of stops is significantly
higher. Moreover, with an elevated maximum speed of \unit[70]{km/h} (Fig.~\ref{fig:complexRun70}), all performance
figures increase significantly reaching up to 
\unit[30]{\%} for the reduction of the number of stops for low demand.
 For higher traffic demand (right columns of the
Figs.~\ref{fig:complexRun50} and~\ref{fig:complexRun70}), the relative performance of the 
TLA decreases for all criteria except for the metrics ``dynamic capacity''.

\section{\label{sec:diss}Discussion and Conclusions}

We have investigated, by means of simulation, a concept of a
traffic-light assistant (TLA) containing three driving strategies to optimize
the approach to and starting from traffic lights:
``economic approach'', ``anticipative start'', and ``flying
start''. The strategies are based on V2X communication: In order to
implement the TLA, equipped vehicles must obtain switching information
of the relevant traffic lights and -- as in the self-controlled signal
strategy of Ref.~\cite{lammer2008SelfControl} -- counting data from a detector at
least \unit[100]{m} upstream of the traffic light. Complex simulations
including all interactions show that, for comparatively
low traffic demand, the TLA is effective. To quantify this, we
introduced relative performance indexes which we consider to be the most
universal approach
to assess penetration effects of individual-vehicle based ITS. For our specific setting
(maximum speed \unit[50]{km/h}, cycle time \unit[60]{s}, green time
\unit[30]{s}), we obtained for a low traffic demand performance indexes of about \unit[15]{\%}
for the number of stops, and about \unit[4]{\%} for most other
metrics. We obtain higher values for higher maximum speeds (increase
by a factor of two for \unit[70]{km/h} instead of \unit[50]{km/h}) and
lower cycle times (increase proportional to the inverse cycle time)
while a higher demand lowers the effect (by a factor of about 0.5 near saturation). While the relative performance
is generally lower than that of the
traffic-adaptive ACC on freeways (about~\unit[25]{\%})~\cite{kesting-acc-roysoc}, the
\emph{individual} advantage kicks in with the first equipped vehicle, in contrast to
traffic-adaptive ACC. While the drivers of the equipped vehicles
benefit most, a smaller effect carries over to the non-equipped
followers since, in
order to avoid a collision, they must adopt at least part of the
driving style of the equipped leader.

\section*{Acknowledgments} 
We kindly acknowledges financial support from the Volks\-wagen~AG
within the German research project KOLINE.
 
\ifCLASSOPTIONcaptionsoff
 \newpage
\fi

\bibliographystyle{ieeetr}

\begin{IEEEbiography}[{\includegraphics[width=1in,height=1.25in,clip,keepaspectratio]{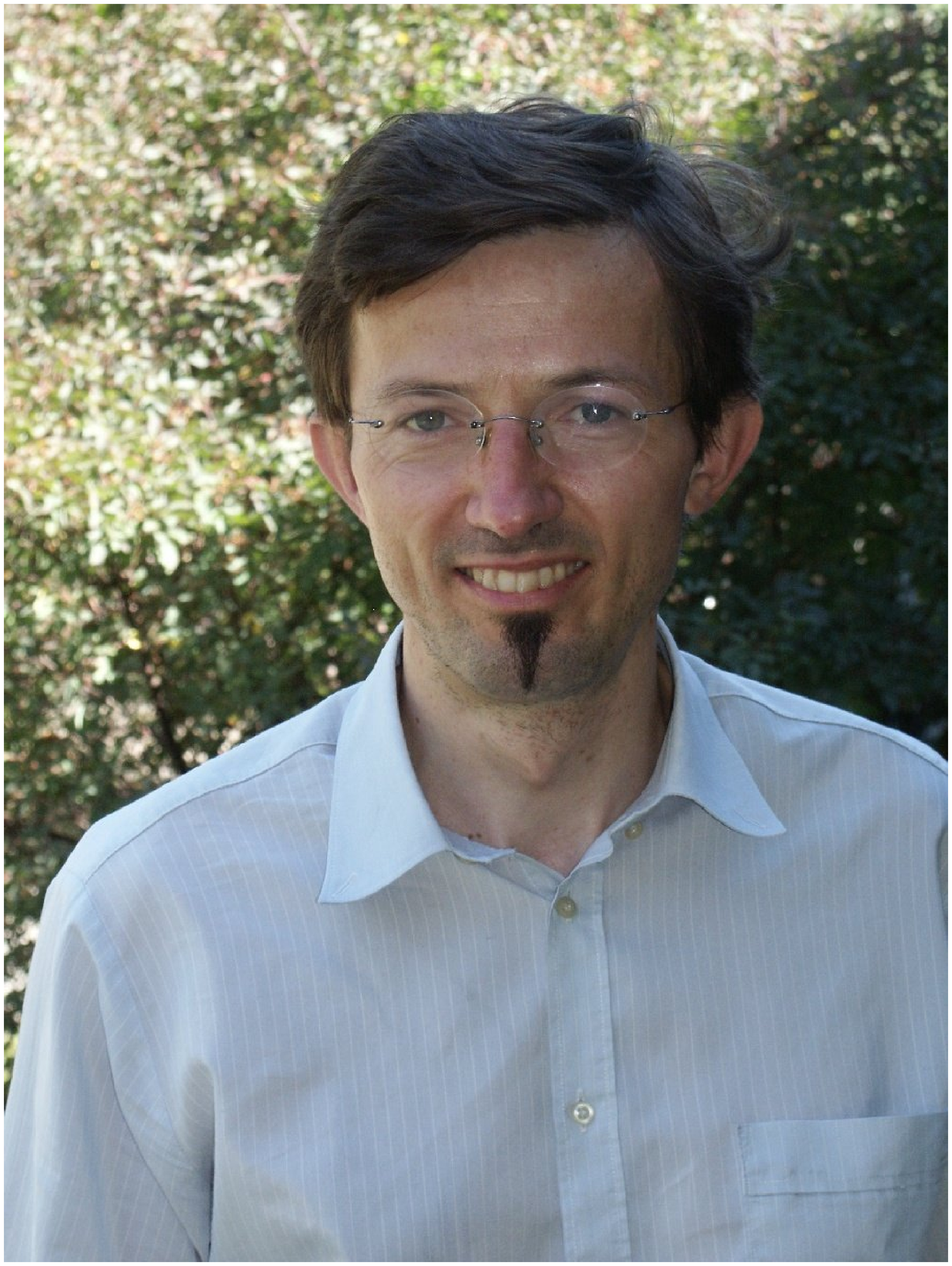}}]{Martin Treiber}
received his Doctoral degree in Physics
in 1996 from Universit\"at Bayreuth, Germany. 
He is lecturer at the Chair for Traffic Modeling and
Econometrics at Technische Universit\"at Dresden, Germany. 
He has been involved in many ITS initiatives and in several research projects of Volkswagen~AG. 
Together with Arne Kesting, he authors the textbook ``Traffic Flow
Dynamics''. His research interests include vehicular
traffic dynamics and modeling, traffic data analysis \& state estimation, driver assistance systems, and the study of macroeconomic impacts of motorized individual traffic. 
\end{IEEEbiography}

\begin{IEEEbiography}[{\includegraphics[width=1in,height=1.25in,clip,keepaspectratio]{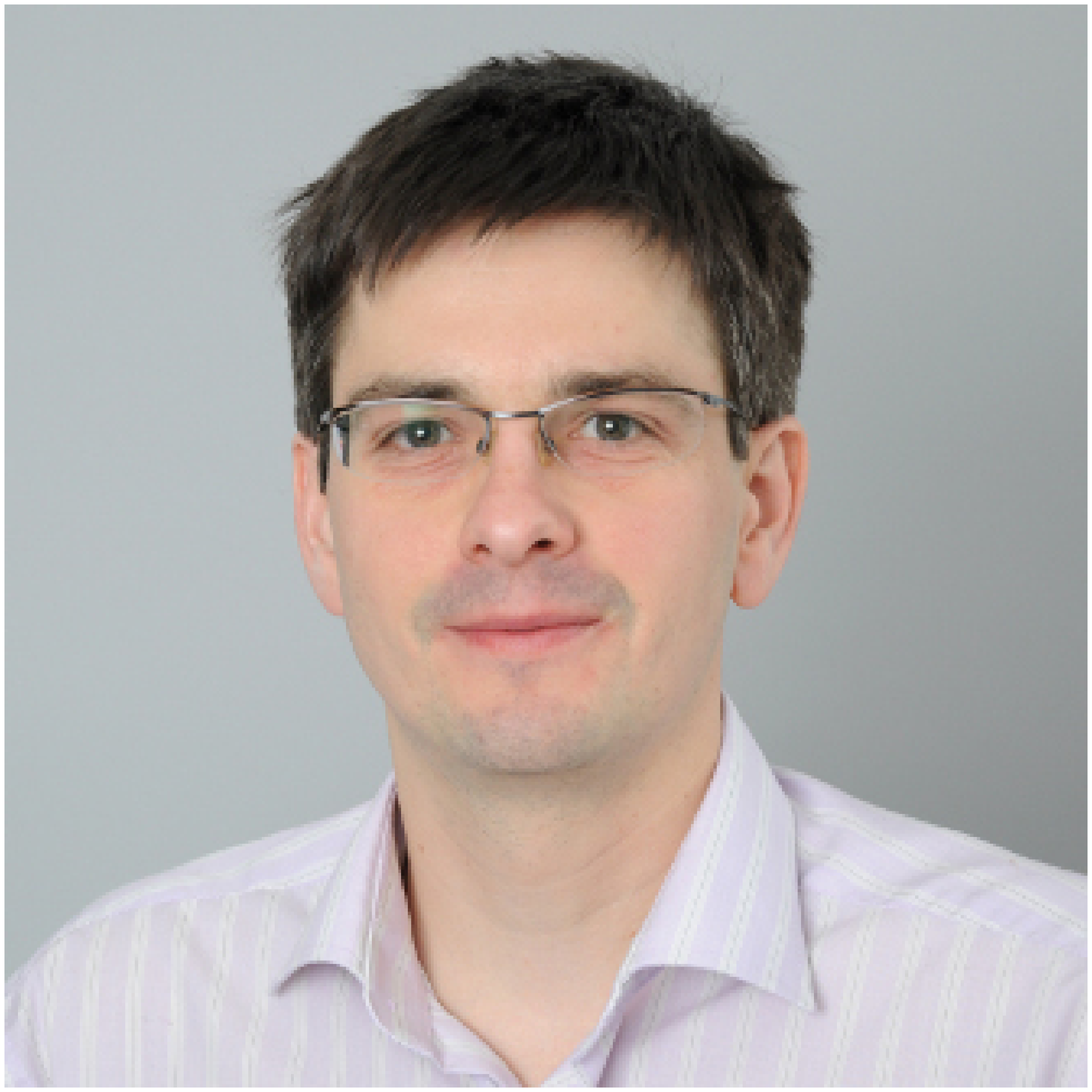}}]{Arne Kesting}
received the Diploma degree in Physics from Freie Universit\"at Berlin, Germany, 
and the Doctoral degree from Technische Universit\"at Dresden, Germany, 
in 2002 and 2008, respectively. In 2009, he received the 
IEEE ITS Best Ph.D. Dissertation Award for the thesis 
entitled ``Microscopic Modeling of Human and Automated Driving: Towards Traffic-Adaptive Cruise Control''. 
He is a senior software developer at TomTom working on real-time traffic information services. 
His research interests include data analysis and fusion, microscopic traffic simulation, and advanced driver-assistant systems.
\end{IEEEbiography}

\end{document}